\documentclass[a4paper,11pt]{article}
\usepackage{axodraw}
\usepackage{graphicx}
\def\euv{\varepsilon_{UV}}
\def\eir{\varepsilon_{IR}}
\def\hyg{{_2F_1}}
\def\s{{\tilde{s}}}
\def\t{{\tilde{t}}}
\def\u{{\tilde{u}}}

\def\m2{{\tilde{m_2^2}}}
\def\sm3{{\tilde{m_3^2}}}

\begin{document}
\begin{flushright}
KEK-CP-121 \\
Nov.~2002
\end{flushright}
\vskip 2cm
\begin{center}
{\Large
QCD event generators with next-to-leading order matrix-elements
and parton showers
}
\end{center}
\vskip 1cm
\begin{center}
{\large Y.~Kurihara$^{\small 1}$, 
J.~Fujimoto$^{\small 1}$, 
T.~Ishikawa$^{\small 1}$, 
K.~Kato$^{\small 2}$, \\
S.~Kawabata$^{\small 1}$,
T.~Munehisa$^{\small 3}$, 
and H.~Tanaka$^{\small 4}$}\\
\vskip 0.5cm
$^{\small 1}${\it High Energy Accelerator Research Organization,\\
Oho 1-1, Tsukuba, Ibaraki 305-0801, Japan}\\
$^{\small 2}${\it Kogakuin University, Shinjiku, Tokyo 163-8677, Japan} \\
$^{\small 3}${\it Yamanashi University, Kofu, Yamanashi 400-8510, Japan} \\
$^{\small 4}$
{\it Rikkyo University, Nishi-Ikebukuro, Toshima, Tokyo 171-8501, 
Japan} 
\end{center}
\vskip 3cm
\begin{abstract}
A new method to construct event-generators based on
next-to-leading order QCD matrix-elements and leading-logarithmic
parton showers is proposed.
Matrix elements of loop diagram as well as those of a tree level
can be generated using an automatic system. A soft/collinear
singularity is treated using a leading-log subtraction method.
Higher order re-summation of the
soft/collinear correction by the parton shower method is combined with
the NLO matrix-element without any double-counting in
this method.

An example of the event generator for Drell-Yan process is given
for demonstrating a validity of this method.
\end{abstract}
\newpage

\section{Introduction}
The standard model(SM) of electromagnetic and strong interactions
has been well established through precision measurements of a large
variety of observables in high-energy experiments\cite{pdg}.
However, there is still one missing part of the SM to be confirmed
by experiments, i.e. the Higgs boson.
In order to search for this missing part and also search for direct signals
beyond the SM,
the LHC experiments\cite{lhc} are under construction at CERN.
Though LHC has employed a proton-proton colliding machine due to
their high discovery potential for new (heavy) particles,
a capability for precision measurements must also be
seriously considered. For extracting physically meaningful information
from a large amount of experimental data, suffered from huge QCD backgrounds,
a deep understanding of the behavior of QCD backgrounds is essentially
important.

The prediction power of a lowest calculation of QCD processes is
very limited due to a large coupling constant of the strong
interaction and an ambiguity of the renormalization energy scale.
Predictions based on the NLO (or more)
matrix elements combined with some all-order summation are
desirable.
Much effort to calculate NNLO corrections\cite{nnlo} is being made
for LHC experiments.
Moreover, in order to  suit
for precision measurements, a simple correction
factor (so-called $K$-factor) for the lowest calculation
is clearly not sufficient.
One must construct event generators including NLO matrix elements
with higher order re-summation even for processes with
multi-particle final states. When one tries to do this task,
one may encounter the following difficulties:
\begin{itemize}
\item{the number of diagram contributing to processes with many particles
in a final state is very large,}
\item{many parton-level processes must be combined to construct
a proton-(anti-)proton collision,}
\item{numerical instability in the treatment of a collinear singularity
may appear,}
\item{a careful treatment of matching between matrix elements
and the re-summation part must be required to avoid double counting, and}
\item{negative weight of events in some part of phase space is not
avoidable.}
\end{itemize}
In order to overcome these difficulties, lots of
works have already been done\cite{nnlo,gens}. In this report,
we propose a new method to construct event-generators with
next-to-leading order QCD matrix-elements and leading-logarithmic
parton showers. Our solution for the above problems
will be given
in this report. 

\section{Matrix elements}
A calculation of the matrix elements of hard-scattering processes
is not a simple task when multiple partons are produced.
At the tree level, the {\tt GRACE} system\cite{grace},
an automatic system for generating Feynman diagram and
a {\tt FORTRAN} source-code to evaluate the amplitude, is established.
The system can in principle treat any number of external particles,
and has been used for up to six fermions\cite{tt}
in the final state within a practical CPU time.
In the {\tt GRACE} system, matrix elements are calculated
numerically using a {\tt CHANEL}\cite{chanel}
library based on a helicity amplitude. {\tt CHANEL} contains
routines to evaluate such things as:
wave-functions/spinors at external states,
interaction vertices, and particle propagator.

Based on this method, we have already published an event generator of
four $b$-quark generation processes at proton-(anti-)proton 
collisions at tree-level named {\tt GR@PPA\_4b}\cite{Tsuno:2002ce},
which include all possible tree-level diagram to create
four $b$-quarks in final state with an interface to PYTHIA\cite{pythia}.

For the loop diagram, effective vertices of two- and three-point
functions have been prepared\cite{guide} and implemented in the system.
At first amplitudes 
of the loop diagram
are evaluated in $d$-dimensional space-time.
The dimensionless coupling constant is introduced as
$g \rightarrow {\tilde \mu}^{d/2-2} g$.
After the loop-momentum integration, $UV$ divergences are
eliminated by subtracting
$1/\euv+\gamma_E-\ln4\pi$ 
where $\euv=2-d/2$ and $\gamma_E$ is the Euler's constant.
This is so-called $\overline{MS}$ 
renormalization scheme\cite{msb}.
Then we are left with the infra-red divergence which is also regularized
using the dimensional regularization by setting
$d=4+2\eir$, where $\eir>0$.


Expressions of effective vertices after the $\overline{MS}$ renormalization
are summarized in Appendix~A.
For more than
three-point diagram, an extended {\tt CHANEL} is prepared. It can evaluate
a fermion current including any number of gamma matrices.
An amplitude of loop diagram will be calculated based on the fermion
current connected by boson propagators. When a fermion loop is
included in the diagram, numerical calculation will be performed after taking
its trace. A bosonic loop can also be treated by the {\tt CHANEL}.

\section{Loop integral}
As two- and three-point functions are embedded in the system as effective 
vertices, loop integrals which must be done in each process start from
four-point functions. The loop-momentum integration of the four-point
diagram can be expressed as
\begin{eqnarray}
I_4(p_1,p_2,p_3,p_4)&=&\left(\mu^2\right)^{\euv} \int  \frac{d^dk}{(2 \pi)^di}
\frac{1}{A_1 A_2 A_3 A_4}, \\
A_1&=&k^2+i0, \nonumber \\
A_2&=&(k+p_1)^2+i0, \nonumber \\
A_3&=&(k+p_1+p_2)^2+i0, \nonumber \\
A_4&=&(k+p_1+p_2+p_3)^2+i0, \nonumber \\
p_1+p_2+p_3+p_4&=&0,\nonumber 
\end{eqnarray}
where $k$ is a loop momentum and $p_i$'s are external momentum taken to
be incoming.
After a momentum integration using an usual procedure 
of Feynman-parameter method, the integral
can be represented as
\begin{eqnarray}
I_4(p_1,p_2,p_3,p_4)&=&\frac{\Gamma(2-\eir)} {(4 \pi)^2}
 {\tilde J}_4(s,t,m_1^2,m_2^2,m_3^2,m_4^2),
\end{eqnarray}
where $m_i^2=p_i^2$, $s=(p_1+p_2)^2$, and $t=(p_1+p_4)^2$.
In general we need tensor integrals
\begin{eqnarray}
&~&J_4(s,t;n_x,n_y,n_z)=  \nonumber \\
&~&\left(4 \pi \mu^2\right)^{-\eir}
\int_0^1 dx~dy~dz
\frac{x^{n_x}y^{n_y}z^{n_z}}
{\left(
-xzs-y(1-x-y-z)t-i0
\right)^{2-\eir}}, 
\end{eqnarray}
here we set all external particles on-shell (massless).
A method to perform this integration is well established.
The scaler integrations (when $n_x=n_y=n_z=0$) without 
an infrared divergence are given in 
\cite{loop1}
and those with IR divergence are given in 
\cite{loop2}.
Tensor integrals can be obtained from a linear
combination of scaler integrals
\cite{loop3}.
Even though the method is well established, we decided to develop 
our own method for the tensor integration of an IR divergent case,
in order to avoid a possible numerical instability due to some
cancellation among IR divergent scaler integrals.

The tensor integration can be done analytically and be represented
by a finite number of terms with Beta and 
Hypergeometric functions as;
\begin{eqnarray}
&~&J_4(s,t;n_x,n_y,n_z)
=\frac{1}{s~t}\frac{B(n_x+\eir,n_y+n_z+\eir)}{1-\eir} \nonumber \\
&\times&\biggl[
\left(\frac{-\t}{4 \pi \mu^2}\right)^{\eir}
\left(\frac{-t}{s}\right)^{n_x}
\frac{n_x}{\prod_{j=1}^{n_x}({n_x}-j+\eir)}B(1+n_z,n_x+n_y+\eir)
\nonumber \\
&\times&\hyg\left(1+n_x,n_x+n_z+\eir,1+n_x+n_y+n_z+\eir,-\frac{\u}{\s}\right)
\nonumber \\
&+&\left(\frac{-\s}{4 \pi \mu^2}\right)^{\eir}
\sum_{l=0}^{n_x} \left(\frac{-s}{t}\right)^l
\frac{\prod_{j=1}^l(l-j-n_x)}{\prod_{j=1}^l(l-j+\eir)}B(1+n_y,l+n_z+\eir)
\nonumber \\
&\times&\hyg\left(1+l,l+n_z+\eir,1+l+n_y+n_z+\eir,-\frac{\u}{\bar t}\right)
\biggr], \\
{\s}&=&s+i0, \nonumber \\
{\t}&=&t+i0,  \nonumber\\
{\bar t}&=&t-i0, \nonumber \\
{\u}&=&u+i0=(p_1+p_3)^2+i0.\nonumber 
\end{eqnarray}
When the numerator is unity ($n_x=n_y=n_z=0$), the scalar
integration can be obtained as
\begin{eqnarray}
&~&J_4(s,t;0,0,0)=\frac{1}{s~t}\frac{B(\eir,\eir)}{\eir(1-\eir)}
\nonumber \\
&\times&\left[\left(\frac{-\s}{4 \pi \mu^2}\right)^{\eir}
\hyg\left(1,\eir,1+\eir,-\frac{\u}{\bar t}\right)
+\left(\frac{-\t}{4 \pi \mu^2}\right)^{\eir}
\hyg\left(1,\eir,1+\eir,-\frac{\u}{\s}\right)\right]
\end{eqnarray}
A explicit structure of the infra-red singularity can be
obtained with Laurent expansion of above formula with respect to
$\eir$.
A full set of formulae and a computer program for the numerical
calculation can be found in \cite{kuri-si}

For more than four-point functions, a reduction formula\cite{guillet}
is used in the system.

%
%
\begin{figure}[b]
\begin{center}
\includegraphics[width=0.5\linewidth]{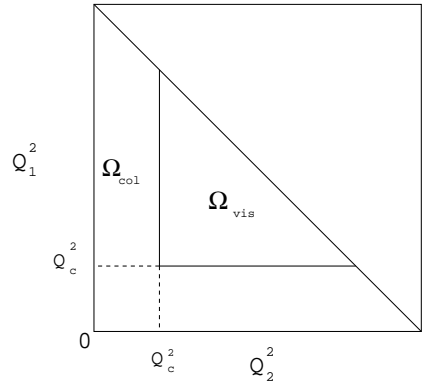}
\caption{\footnotesize{
Dalitz plot of (N+1)-body phase space. $Q^2_i$ is defined in
Eq.(\ref{Ocoll}) and $Q^2_c$ is a threshold value to separate
the visible- and collinear-region.
}}
\label{dplot}
\end{center}
\end{figure}
\section{Real emission}
\subsection{Soft/Collinear treatment}
At first, the conventional method of phase-space slicing\cite{slice} is used
to treat soft and collinear singularities. It is explained for the
case of an initial-state radiation in this report.
The final-state radiation can be
treated in a similar way.
Let's consider two colored partons, whose momenta are $p^\mu_1$ and $p^\mu_2$,
scattered into the $N$-body final state of colorless particles
as a Born process.
For the radiative
correction of this process, we must treat processes with one additional
colored parton, whose momentum is $k^\mu$,
emitted in addition to the Born process.
The matrix elements of the real emission processes must be integrated
in ($N$+1)-body phase-space, $\Phi^{(d)}_{N+1}$, in a $d$-dimensional
space-time.
The space-time dimension is set to be
$d=4+2\varepsilon_{IR}$ in this report.
A collinear region in the ($N+1$)-body phase-space
of the final particles in the $d$-dimensional space-time is defined as
\begin{eqnarray}
\Omega_{coll}&=&\Phi^{(d)}_{N+1} \wedge \{k^\mu | Q^2_i=-(p_i-k)^2<Q^2_c\}
\subset\Omega_{full}=\Phi^{(d)}_{N+1},
\label{Ocoll}
\end{eqnarray}
as shown in figure~\ref{dplot},
where $Q^2_c$ is some cut-off value. It must be sufficiently small
so as not to be
observed experimentally. Final results must be independent of
this value.
In the collinear region, matrix elements can be approximated as
\begin{eqnarray}
\biggl|{\cal M}^{(d)}_{N+1}\biggr|^2&=&
\biggl|{\cal M}^{(4)}_{N}\biggl(q\rightarrow \sum_{i=1}^N q_i\biggr)\biggr|^2
\frac{16\pi}{s \mu^{2\varepsilon_{IR}}}
f_c \frac{\alpha_s}{2\pi} {\tt P}(x)
\frac{1}{k_{T}^2}\biggl(\frac{1-x}{x}\biggr),  \\
q^\mu&=&p_1^\mu+p_2^\mu-k^\mu, \nonumber
\end{eqnarray}
where ${\tt P}(x)$ is a splitting function
for a given parton splitting at a leading-logarithmic order,
$k_T$ is the transverse momentum of the radiated parton, 
$f_c$ is a color factor of a given branching and
$\mu$ the energy scale of the splitting.
The CM-energy of the Born process is
${\hat s}=q^2=s x$, where $s=(p_1+p_2)^2$.
Here, we can set $d$ to be four in the matrix element of
the Born process, since it has no IR singularity.
In the same approximation, the phase space is expressed as
\begin{eqnarray}
d\Phi^{(d)}_{N+1}&=&d\Phi^{(4)}_N\biggl(q\rightarrow \sum_{i=1}^N q_i\biggr)
\nonumber \\
&\times&
\frac{1}{16\pi^2 \Gamma(1+\varepsilon_{IR})}
\biggl(\frac{k_T^{2}}{4 \pi x^2}\biggr)^{\varepsilon_{IR}}
\frac{1}{1-x}
dx dk_T^2.
\end{eqnarray}
The total cross section of real-radiation processes in the collinear region
can be obtained as
\begin{eqnarray}
\sigma_{coll}
&=&\frac{1}{(2p^0_1)(2p^0_2)v_{rel}}
\int_{\Omega_{coll}} d\Phi^{(d)}_{N+1} \biggl|{\cal M}^{(d)}_{N+1}\biggr|^2, \\
&=&\frac{1}{(2p^0_1)(2p^0_2)v_{rel}}
\biggl[\int_{\Omega_{full}}d\Phi^{(d)}_{N+1}-
\int_{\Omega_{vis}}d\Phi^{(d)}_{N+1}\biggr]
\biggl|{\cal M}^{(d)}_{N+1}\biggr|^2,
\end{eqnarray}
where $\Omega_{vis}$ means a visible region of the phase space, such as
\begin{eqnarray}
\Omega_{vis}&=&\Omega_{full}-\Omega_{coll}.
\end{eqnarray}
Then, the collinear cross section can be obtained after the integration
with respect to $k^2_T$ as
\begin{eqnarray}
\sigma_{coll}
&=&\sigma_{full}-\sigma_{vis},  \\
\sigma_{full}&=&\frac{1}{(2p^0_1)(2p^0_2)v_{rel}}
\int_{\Omega_{full}}d\Phi^{(d)}_{N+1}
\biggl|{\cal M}^{(d)}_{N+1}\biggr|^2,  \\
&=&
\left(\frac{s}{4 \pi \mu^2}\right)^{\varepsilon_{IR}}
\frac{B(\varepsilon_{IR},\varepsilon_{IR})}{2\Gamma(1+\varepsilon_{IR})}
f_c \frac{\alpha_s}{2\pi}
\int_0^1 dx  \sigma_{0}(x s)
{\tt P}(x)
\left(\frac{1-x}{x}\right)^{2\varepsilon_{IR}}, \\
\sigma_{vis}&=&\frac{1}{(2p^0_1)(2p^0_2)v_{rel}}
\int_{\Omega_{vis}}d\Phi^{(d)}_{N+1}
\biggl|{\cal M}^{(d)}_{N+1}\biggr|^2,  \\
&=&
2 f_c\frac{\alpha_s}{2\pi} \int_0^1 dx \sigma_{0}(x s)
{\tt P}(x)\ln{\biggl(\frac{s}{Q_c^2}(1-x)-1\biggr)}
\Theta\biggl(1-\frac{2Q_c^2}{s}-x\biggr), 
\end{eqnarray}
where
\begin{eqnarray}
\sigma_0(xs)&=& \frac{1}{x(2p^0_1)(2p^0_2)v_{rel}} \int d\Phi^{(4)}_N
\biggl|{\cal M}^{(4)}_{N}\biggl(q\rightarrow \sum_{i=1}^N q_i\biggr)\biggr
|^2_{q^2=xs}, 
\end{eqnarray}
is the Born cross-section at a CM energy of $q^2=xs$
in four-dimensional
space-time, 
$\mu$ is a factorization energy scale,
and
$\Theta$ is a step function.
For example in case of a splitting for a quark to quark+gluon,
$\sigma_{coll}$ after the $\eir$ expansion becomes 
\begin{eqnarray}
\sigma_{coll}
&=&\sigma_{0}(s) \frac{\alpha_s}{2 \pi} f_{c}
\biggl[\frac{2}{\varepsilon_{IR}^2}+\frac{2L-3}{\varepsilon_{IR}}
-\frac{\pi^2}{2}+L^2\biggr]  \nonumber \\
&+&2\int_0^1 dx \sigma_{0}(x s) \phi(x,\varepsilon_{IR})\nonumber \\
&+&2f_c\frac{\alpha_s}{2\pi} \int_0^1 dx \sigma_{0}(x s) \left[
L\frac{1+x^2}{(1-x)_+}+
2\frac{(1+x^2)\ln{(1-x)}}{(1-x)_+}
-\frac{1+x^2}{1-x}\ln{x}
\right]\nonumber \\
&-&2 f_c\frac{\alpha_s}{2\pi} \int_0^1 dx \sigma_{0}(x s)
\frac{1+x^2}{(1-x)_+}
\ln{\biggl(\frac{s}{Q_c^2}(1-x)-1\biggr)}
\Theta\biggl(1-\frac{2Q_c^2}{s}-x\biggr), 
\end{eqnarray}
where
\begin{eqnarray}
\phi(x,\varepsilon_{IR})&=&
\frac{1}{\varepsilon_{IR}} f_c
\frac{\alpha_s}{2 \pi} {\tt P}(x)=
\frac{1}{\varepsilon_{IR}} f_c
\frac{\alpha_s}{2 \pi}\frac{1+x^2}{(1-x)_+}
, \\
L&=&\ln{(s/\mu^2)}.
\end{eqnarray}
The IR-divergent terms in the collinear cross section can be
canceled out after summing up terms from virtual corrections.
After combining with virtual corrections,
there still exists the collinear-divergent term,
$\phi(x,\varepsilon_{IR})$.
This term is thrown away by hand, because it is counted in the PDF or
PS on the initial partons.
Since the $\overline{MS}$ renormalization scheme is used in this calculation,
the same scheme must be used in the PDF or PS.

%
%
\subsection{Visible jet cross-section}
The last part of the NLO cross-section calculation
is a real emission of the additional parton into the visible region
with exact matrix elements. Those matrix elements of ($N$+1)-parton
production at the final state in four-dimensional space-time
can be automatically generated using
the {\tt GRACE} system.
Cross sections are obtained by integrating those
matrix elements under the four-dimensional phase-space as
\begin{eqnarray}
\sigma_{exact}&=&\frac{1}{(2p^0_1)(2p^0_2)v_{rel}}
\int_{\Omega_{vis}} d\Phi^{(4)}_{N+1}
\biggl|{\cal M}^{(4)}_{N+1}\biggr|^2.
\end{eqnarray}
The space-time dimension is set to be four ($\varepsilon_{IR}=0$) hereafter,
since there is no IR-divergence
in $\Omega_{vis}$.
When the threshold energy ($Q_c^2$) is set to be sufficiently small, this
cross section can be larger than the Born cross section due to
the large magnitude of the coupling constant ($\alpha_s$).
A parton-level calculation has no problem, except for  
the large higher-order correction. However, if one tries to calculate
the cross sections of a proton (anti-)proton collision, one has to combine
the matrix elements with the PDF or PS to construct a proton from
partons. The PDF and PS include leading-log(LL) terms of the initial-state
parton emission. 
If the matrix element is combined with the PDF or PS very naively,
a double-counting of these LL-terms must be unavoidable.
Our proposal to solve this problem is to subtract the LL-terms
from the exact matrix elements as
\begin{eqnarray}
\sigma_{LLsub}&=&\frac{1}{(2p^0_1)(2p^0_2)v_{rel}}
\int_{\Omega_{vis}} d\Phi^{(4)}_{N+1} \biggl[
\biggl|{\cal M}^{(4)}_{N+1}\biggr|^2-
\biggl|{\cal M}^{(4)}_{N}(s x)\biggr|^2 f_{LL}(x,s) \biggr],  \\
f_{LL}(x,s)&=&f_c \frac{\alpha_s}{2 \pi} {\tt P}(x)
\frac{16 \pi^2}{k_T^2}\biggl(\frac{1-x}{x}\biggr).
\end{eqnarray}
The second term of the integrand is the LL-approximation of the
real-emission matrix-elements under the collinear condition.
There is nothing but '$\sigma_{vis}$' in the last subsection.
Then the real-emission cross section ($\sigma_{real}$) can be expressed as
\begin{eqnarray}
\sigma_{real}
&=&\sigma_{col}+\sigma_{exact},\label{re1} \\
&=&\sigma_{full}-\sigma_{vis}+\sigma_{exact}, \\
&=&\sigma_{full}+\sigma_{LLsub}.\label{re2}
\end{eqnarray}
One can expect a numerically stable calculation using Eq.(\ref{re2})
rather than Eq.(\ref{re1}), because cancellation between
$\sigma_{vis}$ and $\sigma_{exact}$ occurred before the phase-space integration.
When the threshold value of $Q^2_c$ is set to be sufficiently small,
the integrand in $\sigma_{LLsub}$ is very close to zero, because the
LL-approximation is very precise around the collinear region. Then,
the result of integration in $\sigma_{LLsub}$ is independent of
the threshold value of $Q_c^2$.

This subtraction may distort the experimental
observables of additional jet distributions. However, a subtracted LL-part
will be recovered after adding the PS applied in the Born process.
The exact matrix element of real-radiation processes gives only
non-logarithmic terms to the visible distributions in the LL-subtraction
method.
The relation between
the PS and the LL-subtraction method is as follows.
The LL-approximation cross section in $\sigma_{LLsub}$ is
\begin{eqnarray}
\sigma_{LL}&=&\frac{1}{(2p^0_1)(2p^0_2)v_{rel}}
\int_{\Omega_{vis}} d\Phi^{(4)}_{N+1}
\biggl|{\cal M}^{(4)}_{N}(s x)\biggr|^2 f_{LL}(x,s) \\
&=&\int_0^1 dx \sigma_0(x s) {\cal D}^{\small (1)}(x),\\
{\cal D}^{\small (1)}(x)&=& \int_{Q_c^2/(1-x)}^{Q_{max}^2}\frac{dk_T^2}
{k_T^2} f_c \frac{\alpha_s}{2 \pi} {\tt P}(x),
\end{eqnarray}
where $Q_{max}^2$ is the maximum value of the transverse momentum.
${\cal D}^{\small (1)}(x)$ is just the first term of the all-order
re-summation of
the leading logarithmic terms in the PS.
Actually, some special case of the Sudakov form factor can be obtained from
${\cal D}^{\small (1)}(x)$as:
\begin{eqnarray}
\Pi(Q^2_{max},Q^2_c/(1-x))&=&
exp\biggl(-\int^{1}_{0} {\cal D}^{\small (1)}(x) dx\biggr).
\end{eqnarray}
However, there is one essential difference between ${\cal D}^{\small (1)}(x)$
and the Sudakov form factor in the PS:
the upper bound of $k_T^2$ integration.
In the PS, this integration is performed up to the energy scale
determined by the Born process instead of their maximum value
in kinetically allowed region.
Then, if
$\sigma_{LL}$ is subtracted in the full phase space of the ($N$+1)-particle
final state, some part of the LL cross section cannot be recovered
from the Born cross-section with the PS. Then, an
appropriate restriction on the phase-space integration must be applied,
such as
\begin{eqnarray}
{\tilde \sigma}_{LLsub}&=&\frac{1}{(2p^0_1)(2p^0_2)v_{rel}}
\int_{\Omega_{vis}} d\Phi^{(4)}_{N+1}
\biggl[
\biggl|{\cal M}^{(4)}_{N+1}\biggr|^2 -
\biggl|{\cal M}^{(4)}_{N}(s x)\biggr|^2 f_{LL}(x,s)
\Theta(Q_{B}^2-k_T^2)
\biggr], \label{LLbar}
\end{eqnarray}
where $Q_{B}^2$ is the energy scale of the Born process, which
must be the same as the factorization energy-scale of the PS.

\begin{figure}[t]
\begin{center}
\includegraphics[width=0.8\linewidth]{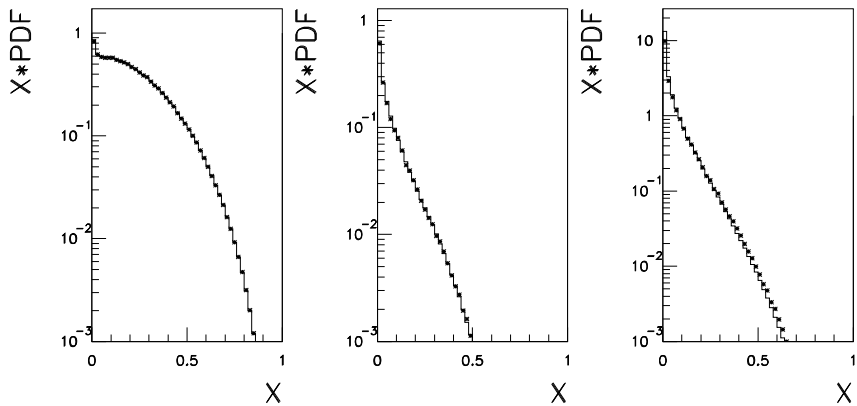}
\caption{\footnotesize{
Momentum fraction distributions from an x-deterministic parton shower
and Cteq5L parton distribution functions. The distributions of
$u$-quarks (left), ${\bar u}$-quarks (center), and
gluons (right) are shown.
}}
\label{qcdps}
\end{center}
\end{figure}
\section{Event generation with a parton shower}
\subsection{Conventional method}
The calculations of cross sections for proton-(anti) proton collisions
are usually performed as follows\cite{pythia,herwig}:
\begin{enumerate}
\item{matrix elements of a given process
at a parton level are prepared,}
\item{a probability to observe a parton in a proton at given energy
scale and momentum fraction is
obtained from some parton distribution function(PDF)\cite{pdf},}
\item{a numerical integration is performed in phase-space
with some initial energy distribution in the PDF,}
\item{for initial-state partons,
a (backward) parton shower\cite{ps}
is applied from the given (high) energy scale to a low energy-scale of
partons,}
\item{for final-state partons, a parton shower is also applied
from a given energy scale to close to the hadron energy-scale,}
\item{some non-perturbative effects according to
string fragmentation or color clustering are taken into account, and}
\item{physical hadrons are formed based on some hadronization model.}
\end{enumerate}
In this method, the parton shower(PS) is used in a supplementary way,
just for generating multi-partons with a finite transverse momentum.

\subsection{An $x$-deterministic PS}
The parton shower
is a method to solve a DGLAP evolution equation\cite{dglap}
using a Monte Carlo method. The PDF's are also obtained by solving
the DGLAP equation with experimental data-fitting.
When an initial distribution of partons is given at some energy scale,
the PS can reproduce a consistent result as the PDF.
For singlet partons, we employ an evolution scheme based on
momentum distributions rather than those on particle-number
distributions, which is used for non-singlet processes.
The momentum-distribution evolution-scheme 
has been proposed by Tanaka and Munehisa\cite{tm} and
shows numerically stable results with high efficiency.

The PDF can give the weight of partons when the momentum fraction ($x$) and
the energy scale are given by the users.
On the other hand, in the forward evolution scheme of the PS,
the momentum fraction of a parton ($x$)
is determined only after evolution takes place. This method is very
inefficient, for instance, for a narrow-resonance production.
In order to cure this inefficiency, we have developed an
$x$-deterministic PS.
In a Monte-Carlo procedure in the PS,
the $Q^2$ evolution is controlled by the Sudakov form-factor, which gives
a non-branching probability of partons,
and the $x$ determination is done independently
according to a splitting function.
It is not necessary to determine the $x$ value at each branching.
After preforming all branching procedures using the Sudakov form-factor,
the $x$ value at each branching is determined to give a total $x$, being
a given value from outside of the PS. In this PS, each event has
a different weight according to the splitting functions and initial
distribution of the PDF.
When the number of branchings is $n$, the weight of this 
event ($W$) can be obtained as
\begin{eqnarray}
W&=&\frac{1}{W_0}\prod_{i=1}^n P(x_i) F_{\small PDF}(x_0) dx_i dx_0,
\label{eq1} \\
W_0&=&\prod_{j=1}^n \int_{0}^{1} P(x_j) dx_j,
\end{eqnarray}
where $P(x_i)$ is a splitting function, $F_{\small PDF}(x_0)$ is
parton distribution function, and $x_0$ is a momentum fraction
before the parton-shower evolution. The momentum fraction after
the evolution is $x=\prod_{i=0}^n x_i$. 
In order to take this momentum fraction ($x$)
as an independent variable given by users, unity 
\begin{eqnarray}
1=\int\delta(x-{\small x_0 \prod_{j=1}^n x_j})dx
\end{eqnarray}
is multiplied
to Eq.(\ref{eq1}) and integration by $dx_0$ is done first as
\begin{eqnarray}
W&=&\frac{1}{W_0}\prod_{i=1}^n P(x_i) F_{\small PDF}(x_0) 
\delta\biggl(x-{\small x_0 \prod_{j=1}^n x_j} \biggr) dx dx_i dx_0, \\
&=&\frac{1}{W_0}\prod_{i=1}^n P(x_i) 
F_{\small PDF}({\tilde x_0}) \frac{dx_i dx}{\prod_{j=1}^n x_j}, \\
{\tilde x_0}&=&\frac{x}{\prod_{j=1}^n x_j}.
\end{eqnarray}
If we
employ an appropriate transformation of the input random numbers,
we can expect a sufficient efficiency.

A numerical test of the $x$-deterministic PS is done by comparing
results with the PDF of {\tt Cteq5L}\cite{cteq}. The initial distributions
of partons are taken from {\tt Cteq5L} at the energy scale of
the $b$-quark mass. Then, after evolution to an energy scale of
100 GeV, the $x$ distributions from the PS are compared to
those in {\tt Cteq5L}.
One can see in Fig.\ref{qcdps} the PS reproduced
$x$ distributions in {\tt Cteq5L} for both of singlet and non-singlet
partons.

Our procedure to generate QCD events involves the following three steps,
\renewcommand{\labelenumi}{\theenumi$'$.}:
\begin{enumerate}
\setcounter{enumi}{1}
\item{numerical integration is performed in the phase-space
including $x$ integration,}
\item{the probability for the existence of each parton
in the proton at some low-energy scale is
obtained from the PDF, and}
\item{the $x$-deterministic PS performs evolution from a low energy scale
to the energy scale of a hard scattering
while generating multi-partons with a finite transverse momentum,}
\end{enumerate}
instead of steps 2, 3, and 4 in the conventional procedure.

\subsection{Negative weight treatment}
Though the LL-subtraction method makes the negative-positive
cancellation in the collinear region very mild, the NLO amplitude
may still have negative values in some region in the phase-space.
This negative weight is unavoidable due to the perturbative
calculation truncated at a finite order of expansion.
Especially this problem is very serious in QCD due to its large
expansion coefficient.
Even if there are events with negative weight, experimentally measurable
distributions can be positive with a realistic detector resolution
(or a bin-width of histograms). Then we decided to employ 
a 'quasi-unweighted' event generation as follows. 

The {\tt BASES}\cite{bases} system can treat integrand with negative values and
prepare a table of a probability density according to the absolute value
of the integrand. The {\tt SPRING}\cite{bases,basesnew} will generate
events based on that table with unit-weight with sign $+1$ or $-1$ 
according to a sign of integrand at given phase point.
This sign will be kept during detector simulations and user analysis
programs. Finally some bins of histograms will be incremented or 
decremented by unit amount due to their sign.

\begin{figure}[t]
\begin{center}
\includegraphics[width=0.8\linewidth]{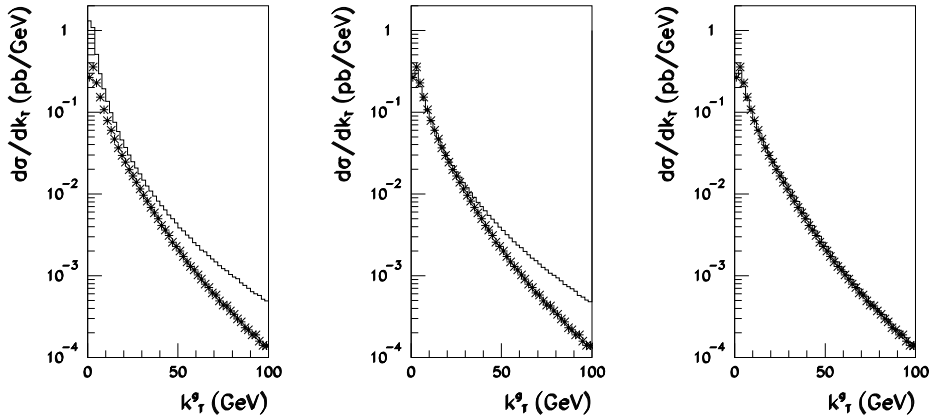}
\caption{\footnotesize{
Transverse momentum distribution of gluons. The distributions
from the PS applied to the Born process are
shown by $*$ commonly in three histograms. Those distributions from
$\sigma_{exact}$ (left), $\sigma_{exact}$ with double-count rejection
(middle), and $\sigma_{exact}$ with double-count rejection and the
$k_T^2$ restriction are compared with the PS.
}}
\label{fig1}
\end{center}
\end{figure}
%
%
\section{A test of the LL-subtraction method}
The LL-subtraction method is numerically tested for a process of
$u {\bar u} \rightarrow \mu^+ \mu^-$ in proton-anti proton collision
at the CM energy of 2 TeV. Here only non-singlet $u$-quark
is used in this test to avoid additional complex problem. Matrix elements
of the Born process and the real radiation process
($u {\bar u} \rightarrow \mu^+ \mu^- gluon$) are generated by {\tt GRACE}.
Numerical integration is done using a {\tt BASES} system.
A non-singlet $u$-quark distribution at the energy scale of
4.6 GeV is taken from {\tt CTEQ5L}. The parton evolution
from this energy scale to that of hard-scattering, i.e. an invariant mass
of the muon-pair (${\hat s}_{\mu\mu}$), is done using the $x$-deterministic PS
for both of the Born and radiative processes. An additional cut of
$\sqrt{{\hat s}_{\mu\mu}}>40$ GeV is also applied.

The distributions of transverse momenta of gluons
are shown in Figure~2. For the Born process, the largest $k_T$ from
the PS is filled to histograms if it has $k_T$ greater than 1 GeV.
For the radiative process, that from matrix elements is filled
when it has passed the same cut as mentioned above. If one does not
take care of double-counting of the LL-terms in the matrix elements
and the PS, the distributions of $d\sigma_{exact}/dk_T$ show
larger values compared with the PS in the Born process, as shown in the
left histogram of Figure~2.
In order to avoid double-counting, we required that
$k_T$ of the matrix elements be greater than those of any gluons
from the PS in the radiative process. This double-counting
rejection leads to a good agreement of the $k_T$ distribution
around the low-$k_T$ region, as shown in the middle histogram.
However, the PS still emits an insufficient
number of gluons around a very high-$k_T$ region.
If the upper bound of the $k_T^2$ integration is set to be
${\hat s}_{\mu\mu}$ in the calculation with the exact matrix elements, 
$k_T^2$ distributions shows a good agreement, as shown in right
histogram. This behavior tells us that the parton-shower can reproduce
gluon distributions only for the energy scale less than ${\hat s}_{\mu\mu}$.
This result confirmed the necessity of the step function
($\Theta(Q^2_B-k_T)$) in Eq.(\ref{LLbar}).

The distributions from ${\tilde \sigma}_{LLsub}$ combined with
those of the PS on the Born process are compared with
those from exact matrix-elements with double-counting rejection.
In both of the gluon energy and the transverse-momentum distributions,
the LL-subtraction method can give consistent results with
those of the exact matrix elements, as shown in Figure~3.
\begin{figure}[t]
\centerline{
\includegraphics[width=0.7\linewidth]{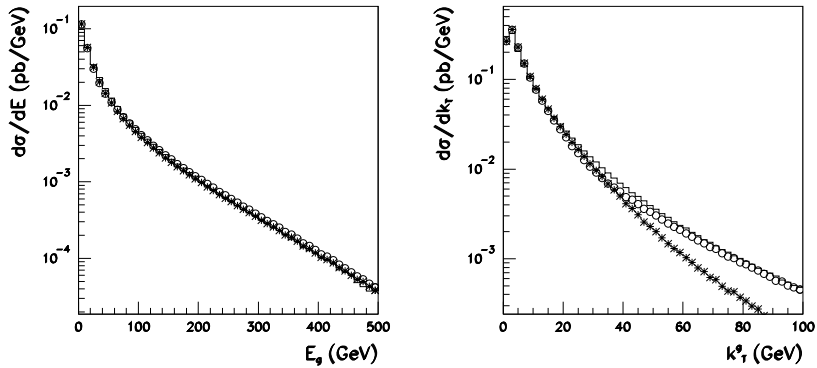}
}
\caption{\footnotesize{
Energy and transverse momentum distributions of gluons.
The $*$ shows the distributions from the PS on the Born process,
solid histograms from $\sigma_{exact}$ with double-count rejection,
and circles from ${\tilde \sigma}_{LLsub}$ combined with those
from the PS on the Born process.
}}
\label{fig2}
\end{figure}
\begin{figure}[t]
\begin{center}
\includegraphics[width=4cm]{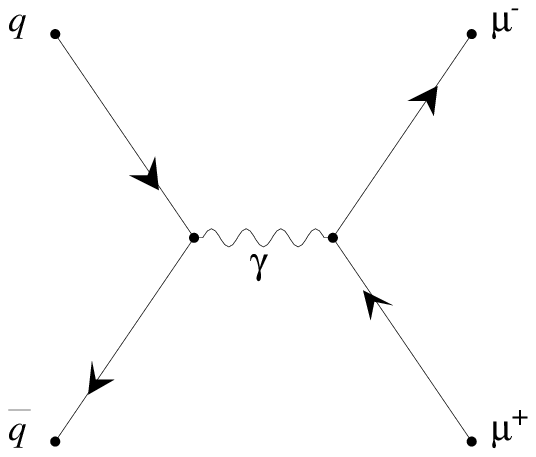}
\includegraphics[width=8cm]{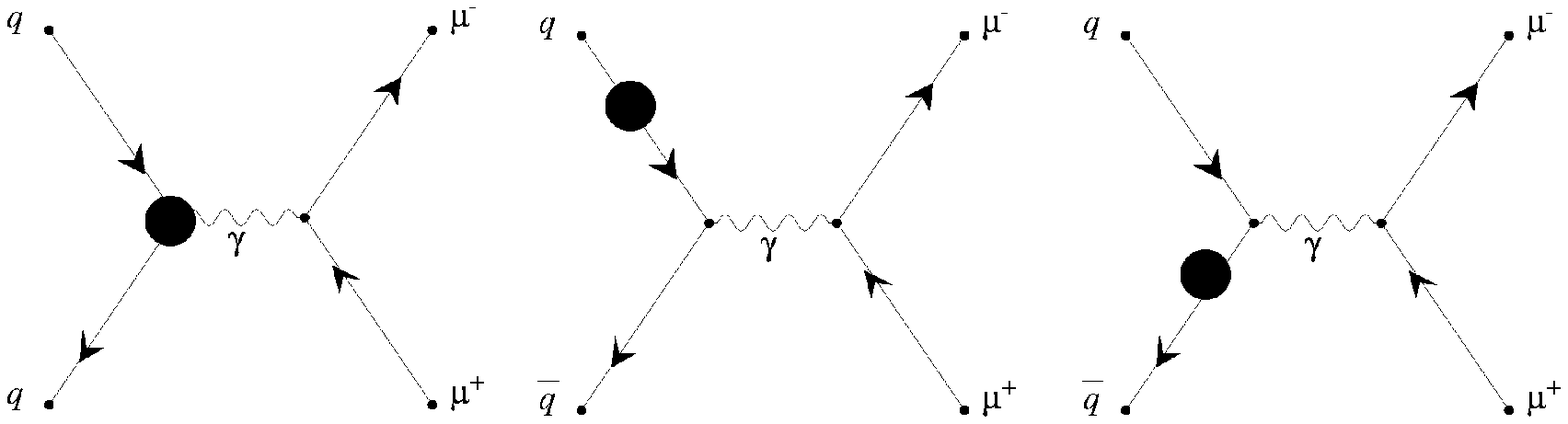}
\includegraphics[width=7cm]{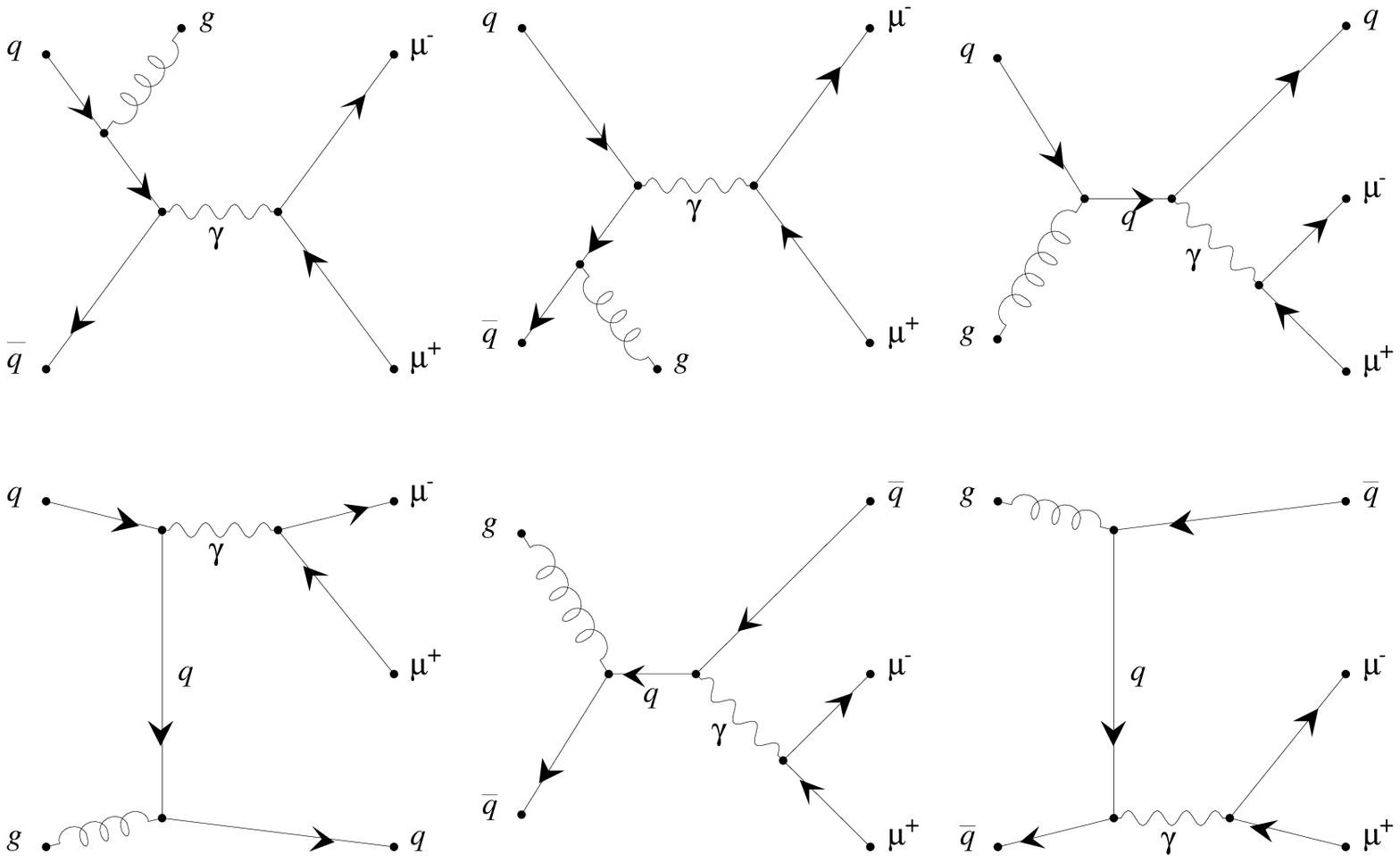}
\caption{\footnotesize{
Feynman diagram of the Drell-Yan process, 
$q {\bar q} \rightarrow {\mu}^- {\mu}^+$ and its radiative processes.
}}
\label{diag}.
\end{center}
\end{figure}
\begin{figure}[t]
\centerline{
\includegraphics[width=10cm]{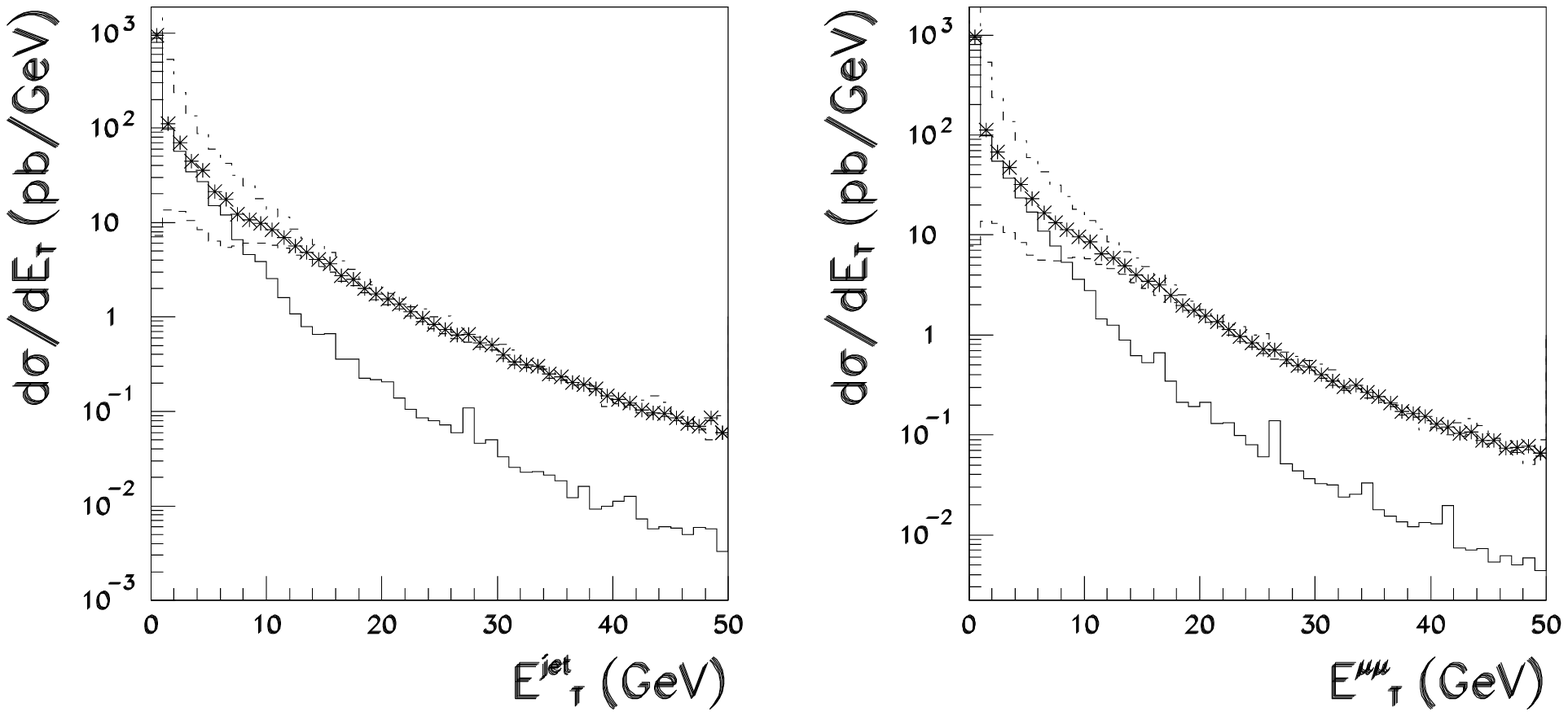}
}
\caption{\footnotesize{
Transverse energy distributions ($E_T$) of jets (left figure) and
muon-pair system (right figure). For a jet distribution,
the largest $E_T$ among jets from parton showers
or matrix elements are filled in the histogram.
Solid histograms show distributions from parton showers applied on
the {\tt tree+virtual+collinear collections}, dashed ones come from
real radiation matrix-elements with PS+(LL-subtraction),
and dot-dashed ones come from the matrix-elements with PDF
without any care of the double-counting.
}}
\label{ptdist}.
\end{figure}
\section{Example:Drell-Yan processes}
In order to demonstrate how our proposed method can really work
for physical processes in hadron colliders, an example of an
event generator for the Drell-Yan process in a proton anti-proton
($p{\bar p}$)
collider of TEVATRON at the next-to-leading order QCD is constructed.
All possible Feynman diagram contributing to the process
$p {\bar p} \rightarrow \mu^+ \mu^-$(+additional parton)
as shown in figure~5 are taken into account in the
generator. All quarks and gluons except top-quark in an initial
(anti-)proton are summed up using the PDF of {\tt Cteq5L} 
at $b$-quark mass of 4.5 GeV and $Q^2$ evolution to the energy
scale of a hard scattering has been done using the $x$-deterministic
parton shower at leading logarithmic order. 

FORTRAN programs to calculate matrix-elements are automatically generated
using GRACE including loop- and real-radiation diagram
as well as tree ones. All quark masses are set to be massless
in the matrix-element calculations. A QCD coupling constant
is calculated using a leading-order formula with $\Lambda_{QCD}=146$ MeV
which is obtained by parameter fittings in {\tt Cteq5L}.
The factorization scale is set to be
equal to the energy of a $\mu$-pair system.
As it is known well for the Drell-Yan process,
the renormalization energy scale does not explicitly appear in
the correction terms. Main correction terms are absorbed
in the running coupling constant and $Q^2$ evolution
of the parton distribution. Remaining terms with the large
logarithm ($L=\ln(Q^2/\mu)Z$) is canceled out after
summing up virtual and collinear correction terms.
This renormalization scale independence is numerically
checked in our program.
Infrared divergent part proportional to $1/\eir^2$ and $1/\eir$ are
left in the program and numerically confirmed that these divergent
coefficients are canceled out after summing up virtual and collinear
parts of the matrix elements at the order of $10^{-20}$. 
No experimental cut except the energy of a $\mu$-pair system
to be greater than 10 GeV is applied.
The total cross section of the Drell-Yan process at 2 TeV of the $p{\bar p}$
CM energy with above cut 
is obtained to be $1.026 \times 10^{3}$ pb at the tree level
and $1.288\times 10^{3}$ pb at the NLO, which gives a $K$-factor of 1.256.
In the cross section of $1.288\times 10^{3}$ pb at the NLO,
$1.807 \times 10^2$ pb comes from the visible cross section after the
LL-subtraction. It is rather good behavior as 
perturbative correction of QCD.
Numerical results of the virtual and collinear correction 
of $q{\bar q}$ initial state are compared with those based on formulae
given by Altarelli et al.\cite{dy-nlo} and are confirmed consistent.
It is confirmed that the cross sections are independent from the value
of $Q^2_c$ defined in section 4.2 and agree with those
with simple phase-space slicing method as expected.

Next let us look at transverse energy distributions of the largest $E_T$
jet and a $\mu$-pair system. Origin of the transverse
energy has two sources, one is finite transverse momentum of 
the PS and the other is that from real-radiation matrix-elements.
In the LL-subtraction method, these two kinds of
distributions can be smoothly combined without any double-counting.
As shown in figure~6, contribution from the parton shower applied
on the NLO matrix elements (solid histograms) 
is dominated around small $E_T$ region.
On the other hand a high $E_T$ events are dominated by 
the real-radiation correction (dashed histograms). 
These two regions are smoothly connected as shown by stars (*)
which show the sum of solid and dashed histograms.
If one applied PDF simply on the the real-radiation
matrix-elements without any care of double counting,
small $E_T$ events are clearly over estimated 
as shown by dot-dashed histograms.
On the other hand, high $E_T$ events are consistent with
those from LL-subtraction method, because they are 
free from both double-counting
and smearing due to the PS. 

\section{Conclusions}
A new method to construct event-generators based on
next-to-leading order QCD matrix-elements and an $x$-deterministic
parton shower is proposed.
Matrix elements of loop diagram as well as those of a tree level
can be generated by the {\tt GRACE} system. A soft/collinear
singularity is treated using the leading-log subtraction method.
It has been demonstrated that the LL-subtraction method can give
good agreement with the exact matrix elements without any
double-counting problem.
The NLO event generator for the Drell-Yan processes has been 
constructed based on our method and showed smooth distributions
of transverse momenta combining the PS and exact matrix-elements
of real radiation processes without any double-counting problem.

\vskip 1cm
Authors would like to thank Dr. S.~Idalia for continuous discussions
on this subject with him and his useful suggestions.

This work was supported in part by the Ministry of Education, Science
and Culture under the Grant-in-Aid No. 11206203 and 14340081.

\newpage

\appendix{{\Huge Appendix}} 
\section{Effective vertices}
\subsection{$SU(N_c)$ Color factor}
\begin{eqnarray*}
{\mathbf T}^a &:&{\rm the~fundamental~representation}, \\
f_{abc} &:&{\rm the~structure~constants}, \\
\biggl[{\mathbf T}^a,{\mathbf T}^b\biggr]&=&i f_{abc} {\mathbf T}^c, \\
\sum_a{\mathbf T}^a{\mathbf T}^a&=&C_F \cdot I 
= \frac{N_c^2-1}{2N_C} \cdot I \\
Tr({\mathbf T}^a{\mathbf T}^b)&=&T_R \delta^{ab}, \\
\sum_{b,c}f_{abc} f_{bcd}&=&C_G \delta^{ad}=N_c\delta^{ad}.
\end{eqnarray*}

\subsection{tree vertex}
\begin{itemize}
\item quark-gluon~vertex: $g{\mathbf T}^a$
\item three-gluon~vertex: $-igf_{abc}$
\item four-gluon~~vertex: $-g^2(f_{abe}f_{cde}+{\rm cyclic~permutation})$
\end{itemize}

\subsection{quark self-energy}
\begin{picture}(300,100)(0,0)
\ArrowLine(80,50)(130,50)
\GCirc(150,50){20}{0.5}
\ArrowLine(170,50)(220,50)
\put(100,30){\makebox(0,0)[rc]{$p~~\rightarrow$}}
\put(250,50){\makebox(0,0)[lc]{{\Large $=\Sigma(p^2)$}}}
\end{picture}
\begin{table}[h]
\begin{center}
\begin{tabular}{|c|l|}
\hline
on-/off-shell & self-energy \\ \hline
~&~ \\
$p^2\ne0$ & $\Sigma(p^2)=C_F \frac{\alpha_s}{4 \pi} (1-\ln\frac{-p^2}
{\mu^2})$ \\ 
~&~ \\
$p^2=0$ & $\Sigma(0)=C_F \frac{\alpha_s}{4 \pi} \frac{1}{\eir}$ \\
~&~ \\
\hline
\end{tabular}
\end{center}
\end{table}

\subsection{gluon self-energy}
\begin{picture}(300,100)(0,0)
\Gluon(80,50)(130,50){5}{4}
\GCirc(150,50){20}{0.5}
\Gluon(170,50)(220,50){5}{4}
\put(100,30){\makebox(0,0)[rc]{$p~~\rightarrow$}}
\put(250,50){\makebox(0,0)[lc]{{\Large $=\Pi_{G}(p^2)$}}}
\end{picture}
\begin{table}[h]
\begin{center}
\begin{tabular}{|c|l|}
\hline
on-/off-shell & self-energy \\ \hline
~&~ \\
$p^2\ne0$ & 
$\Pi_G(p^2)=-C_F \frac{\alpha_s}{4 \pi}(\frac{31}{9}-\frac{5}{3}
\ln\frac{-p^2}{\mu^2})
+T_R N_f \frac{\alpha_s}{4 \pi} 
(\frac{20}{9}-\frac{4}{3}\ln\frac{-p^2}{\mu^2})$ \\
~&~ \\
$p^2=0$ & $\Pi_G(0)=-C_F \frac{\alpha_s}{4 \pi}\frac{1}{\eir}
+T_R N_f \frac{\alpha_s}{4 \pi} \frac{4}{3}\frac{1}{\eir}$ \\
~&~ \\
\hline
\end{tabular}
\end{center}
\end{table}

\subsection{quark-gluon vertex}
\begin{picture}(300,100)(0,0)
\ArrowLine(80,30)(130,30)
\GCirc(150,30){20}{0.5}
\ArrowLine(170,30)(220,30)
\Gluon(150,50)(150,90){5}{4}
\put(100,10){\makebox(0,0)[rc]{$p_1~~\rightarrow$}}
\put(230,10){\makebox(0,0)[rc]{$\leftarrow~~p_2$}}
\put(190,80){\makebox(0,0)[rc]{$\downarrow k^{\mu}$}}
\put(250,50){\makebox(0,0)[lc]{{\Large $=\Lambda_{\mu}^a(p_1,p_2,k)$}}}
\put(150,100){\makebox(0,0)[rc]{$a$}}
\end{picture}
\begin{eqnarray*}
\Lambda_{\mu}(p_1,p_2,k)&=&g {\mathbf T}^a (C_F-\frac{1}{2}C_G)
\frac{\alpha_s}{4 \pi} 
({\cal F}_1^I \gamma^{\mu}
+{\cal F}_2^I\frac{p_j^\mu\not{k}}{-p_j^2}) \\
&+&g {\mathbf T}^a \frac{1}{2}C_G \frac{\alpha_s}{4 \pi}
({\cal F}_1^I \gamma^{\mu}
+{\cal F}_2^I\frac{p_j^\mu\not{k}}{-p_j^2}) 
\end{eqnarray*}
\subsubsection{case I}
$k^2 = 0$, $p_i^2 = 0$, $p_j^2=q^2\ne0$, and 
$L=\ln\frac{-q^2}{\mu^2}$,
where $(i,j)=(1,2)$ or $(i,j)=(2,1)$.
\begin{table}[h]
\begin{center}
\begin{tabular}{|c||l|}
\hline
~&~ \\
${\cal F}^I_1$ & $\frac{2}{\eir}+L-4$ \\
~&~ \\
${\cal F}^I_2$ & $\frac{4}{\eir}+4L-10$ \\
~&~ \\
${\cal F}^{II}_1$ & $-\frac{2}{\eir^2}+\frac{3-2L}{\eir}
+\frac{\pi^2}{6}-L$ \\
~&~ \\
${\cal F}^{II}_2$ & $-\frac{2}{\eir^2}-\frac{2L}{\eir}+
\frac{12+\pi^2-6L^2}{6}$ \\
~&~ \\
\hline 
\end{tabular}
\end{center}
\end{table}
\subsubsection{case II}
$k^2 = q^2$, $p_1^2 = p_2^2 = 0$, and
$L=\ln\frac{-q^2}{\mu^2}$.
\begin{table}[h]
\begin{center}
\begin{tabular}{|c||l|}
\hline
~&~ \\
${\cal F}^I_1$ & $-\frac{2}{\eir^2}-\frac{2L-4}{\eir}-8
+\frac{\pi^2}{6}+3L-L^2$ \\
~&~ \\
${\cal F}^I_2$ & 0 \\
~&~ \\
${\cal F}^{II}_1$ & $\frac{4}{\eir}-2+L$ \\ 
~&~ \\
${\cal F}^{II}_2$ & 0 \\
~&~ \\
\hline 
\end{tabular}
\end{center}
\end{table}

\subsection{three gluon vertex}
\begin{picture}(300,100)(0,0)
\Gluon(80,30)(130,30){5}{4}
\GCirc(150,30){20}{0.5}
\Gluon(170,30)(220,30){5}{4}
\Gluon(150,50)(150,90){5}{4}
\put(100,10){\makebox(0,0)[rc]{$k^{\mu_1}_1~~\rightarrow$}}
\put(230,10){\makebox(0,0)[rc]{$\leftarrow~~k^{\mu_2}_2$}}
\put(190,80){\makebox(0,0)[rc]{$\downarrow k^{\mu_3}_3$}}
\put(70,30){\makebox(0,0)[rc]{$a_1$}}
\put(240,30){\makebox(0,0)[rc]{$a_2$}}
\put(150,100){\makebox(0,0)[rc]{$a_3$}}
\put(250,50){\makebox(0,0)[lc]
{{\Large $=\Lambda_{\mu_1 \mu_2 \mu_3}^{a_1 a_2 a_3}
(k_1,k_2,k_3)$}}}
\end{picture}
\[
k_1^2=k_2^2=0, k_3^2=q^2\ne0, \\
L=\ln\frac{-q^2}{\mu^2}
\]
\begin{eqnarray}
\Lambda_{\mu_1 \mu_2 \mu_3}^{a_1 a_2 a_3} (k_1,k_2,k_3)
&=&
-i g f^{a_1 a_2 a_3} \frac{\alpha_s}{4 \pi}
\biggl[\frac{C_G}{2}\biggl(
{\cal G}_1+
{\cal G}_2+
{\cal G}_3\biggr)
+\frac{N_f}{2}{\cal G}_4 \biggr], \\
{\cal G}_i&=&\sum_{j=1}^3 c_{ij} {\cal P}_j
\end{eqnarray}
\begin{center}
${\cal G}_1$:gluon loop, ${\cal G}_2$:ghost loop, ${\cal G}_3$:gluon loop
(fish type), ${\cal G}_4$:quark loop
\end{center}
{\bf Independent momenta}
\begin{itemize}
\item ${\cal P}_1^{\mu_1 \mu_2 \mu_3}=(k_1-k_2)^{\mu_3} g^{\mu_1 \mu_2}$
\item ${\cal P}_2^{\mu_1 \mu_2 \mu_3}=k_1^{\mu_2} g^{\mu_1 \mu_3}
-k_2^{\mu_2} g^{\mu_2 \mu_3}$
\item ${\cal P}_2^{\mu_1 \mu_2 \mu_3}=\frac{k_2^{\mu_1}k_1^{\mu_2}
(k_1-k_2)^{\mu_3}}{q^2}$
\end{itemize}
\begin{table}[t]
\begin{center}
\begin{tabular}{|c||l|l|l|}
\hline
$c_{ij}$&$j=1$&$j=2$&$j=3$ \\ \hline \hline
&&&\\
$i=1$ &
$
-\frac{3}{2\eir^2}-\frac{3}{2\eir}(-5+L) 
$ & $
\frac{2}{\eir^2}-\frac{1}{2\eir}(19-4L)
$ ~&  \\ ~& $
-\frac{1}{12}(103-51L+9L^2)+\frac{\pi^2}{8}
$ & $
+\frac{1}{3}(19-9L+3L^2)-\frac{\pi^2}{6}
$ & $-\frac{3}{2}$ \\&&& \\
$i=2$ & $
-\frac{11-3L}{36} 
$ & $
\frac{8-3L}{18} 
$ & $
\frac{1}{6} 
$ \\&&& \\
$i=3$ &
$
-\frac{9}{2\eir}
$ & $
\frac{2}{2\eir}+\frac{18-19L}{2}
$ & 
0 \\&&& \\
$i=4$ &
$
\frac{14-12L}{9} 
$&$
-\frac{40-24}{9} 
$&$
\frac{4}{3}
$
\\&&&\\
\hline 
\end{tabular}
\end{center}
\end{table}

\begin{thebibliography}{}
\bibitem{basesnew}Negative weight treatment in {\tt SPRING}
is prepared by S.~Kawabata. It is not implemented in
published version in ref.\cite{bases}. If you need the version
with negative weight treatment, please contant with authors.
\end{thebibliography}
\end{document}